\documentclass[12pt]{article}

\usepackage{amsmath,amsfonts,epsfig,color,latexsym}

\newcommand{\be}{\begin{equation}}
\newcommand{\ee}{\end{equation}}
\newcommand{\bea}{\begin{eqnarray}}
\newcommand{\eea}{\end{eqnarray}}

\let\newsection=\section
\renewcommand{\section}{\setcounter{equation}{0}\newsection}

\begin{document}

\begin{flushright}
hep-th/0703034\\
TIT/HEP-566
\end{flushright}
\vskip.5in

\begin{center}

{\LARGE\bf D=4 Einstein gravity from higher D CS and BI gravity and an 
alternative to dimensional reduction }
\vskip 1in
\centerline{\Large Horatiu Nastase\footnote{email: nastase@phys.titech.ac.jp}}
\vskip .5in

\end{center}
\centerline{\large Global Edge Institute,}
\centerline{\large Tokyo Institute of Technology}
\centerline{\large Tokyo 152-8550, Japan}

\vskip 1in

\begin{abstract}

{\large An alternative to usual dimensional reduction for gravity is analyzed, in the 
vielbein-spin connection formulation. Usual 4d Einstein gravity plus a topological term
(the "Born-Infeld" Lagrangian for gravity), is shown to be obtained by a generalized 
dimensional reduction from 5d Chern-Simons gravity. Chern-Simons gravity in d=2n+1 is 
dimensionally reduced to CS gravity in d=2n-1 via a mechanism similar to descent equations.
The consistency of the dimensional reduction in both cases is analyzed.
The dimensional reduction of d=2n+2 Born-Infeld gravity to d=2n BI gravity, as well as 
d=2n BI gravity to d=2n-1 CS gravity is hard to achieve. Thus 4d gravity (plus a topological
term) can be embedded into d=2n+1 CS gravity, including 11d CS, whose supersymmetric version could possibly be related to usual 11d supergravity. This raises the hope that maybe
4d quantum Einstein gravity could be embedded in a well defined quantum theory, similar
to Witten's treatment of 3d quantum Einstein gravity as a CS theory.  
}

\end{abstract}

\newpage

\section{ Introduction}

Dimensional reduction as a tool for quantum field theory started with the idea of Kaluza 
and Klein to use 5d Einstein gravity (in the usual, metric, formulation) and obtain unification in 4d of gravity with electromagnetism. The inverse process ("dimensional
oxidation") has been used to simplify quantum field theories and supergravity theories 
alike, for example showing that the complicated N=8 supergravity in 4d can be embedded into
the simple and unique N=1 supergravity in 11d. When one does this however, the quantum 
structure of the theory gets more problematic (see for example \cite{bdr,grv} showing that 
although 4d N=8 supergravity shows evidence of fewer or possibly no divergencies, 11d 
supergravity has a more complicated divergence structure). In supergravity, the 
gravity formulation in terms of vielbein or spin connection is more fundamental than the 
metric one (in particular in one of the first, second, or 1.5 order formulations 
\cite{peter}). Yet still, when one talks about dimensional reduction, one adopts a 
formulation that mimics the case of the metric formulation of gravity: find a background
with a geometrical interpretation (e.g. torus, or sphere), expand around it in spherical 
harmonics, and keep only the lowest mass multiplet in the expansion. 

However, this need not a priori be the case. In a famous paper \cite{witten}, 
following earlier work in \cite{pvn,at}, 
Witten showed that following the metric version of the theory too closely one can 
miss important facts. In particular, he noticed that 3d gravity is a Chern-Simons theory,
thus a gauge theory for the Poincare group, with gauge fields $e_{\mu}^a$ and $\omega_{\mu}
^{ab}$. As such, it can be quantized by treating it as a theory on an abstract space, 
with the vielbein and spin connection being just regular fields, with the natural background
value of zero, instead of the natural background value of $e_{\mu}^a=\delta_{\mu}^a, 
\omega_{\mu}^{ab}=0$ of flat space, borrowed from the metric formulation. Although the 
3d Einstein gravity is a gauge theory of CS type, the 4d Einstein gravity is not (the 
Lagrangian is not gauge invariant, or rather it is only gauge invariant on shell, and by 
identifying the base manifold with the tangent space, i.e. diffeomorphisms with gauge 
transformations).

In this paper, I will try to analyze the case of 4d gravity in the vielbein-spin connection formulation, and embed it in higher dimensions, similarly without preconceived notions 
about how dimensional reduction should look like. We will see that in fact one can now
improve the behaviour of the 4d quantum gravity by adding a topological term  and
embedding the resulting Lagrangian ("Born-Infeld" gravity) in 2n+1 dimensional
Chern-Simons gravity theories, which themselves can be embedded in 2n+2 dimensional 
topological theories. While I will not attempt to define the quantum version of the 
Chern-Simons gravities, or to see how it relates to the 4d quantum gravity, it is conceivable that such a treatment will be possible along the lines of Witten's analysis.
After this paper was finished, I became aware of \cite{dm}, which deals with the dimensional
reduction of Lovelock-type actions, but from a different point of view (and with different
aims and results) as this paper, and the related work \cite{mh} which deals with some 
Chern-Simons gravity theories and their relation to Lovelock gravity in odd dimensions. 

I will start in section 2 with an embedding that I had already argued for in 
\cite{nastase}. 
Specifically, I will argue that Chern-Simons gravity in d=2n+1 can be reduced to Born-Infeld
gravity in d=2n, via a natural extension of the usual dimensional reduction of the 
Einstein-Hilbert action. 
I will analyze the dimensional reduction in more detail, and specialize 
for the case of n=2 (d=4 gravity, of the type of Einstein action plus a topological term).
In section 3 I will show that it is possible to go from  CS gravity in d=2n+1 to CS gravity 
in d=2n-1, via a generalized notion of dimensional reduction, with the 
2n+1 to 2n reduction not having a geometric analog, while from 2n to 2n-1  one just goes to 
the boundary theory, the whole sequence being a generalized version of descent equations. 
In section 4 I will attempt to do go from d=2n+2 Born-Infeld gravity to d=2n BI gravity
and from BI to CS gravity, 
and I will see that it is not easy, but I will write down conditions that if satisfied, 
will lead to a successful dimensional reduction. In section 5 I will discuss about 
the possible relations to 11d supergravity and topological theories, and conclude.

\section{From 5d Chern-Simons to 4d Einstein gravity}

In the vielbein-spin connection formulation of gravity, the gravitational action is 
written in a form mimicking gauge theories, via the curvature 2-form
\be
R^{ab}(\omega)= d\omega^{ab}+\omega^{ac}\wedge \omega^{cb}
\ee
The Einstein action is then written as a gauge theory action with the gauge group 
being the Poincare group ISO(d-1,1) and the gauge fields $\omega^{ab}$ (for $J^{ab}$)
and $e_{\mu}^a$ (for $P^a$). In 4d the Einstein action is 
\be
S_{EH}=\int d^4x \epsilon_{abcd} R^{ab}(\omega)\wedge e^c\wedge e^d
\ee
in a first order formulation, since by varying with respect to $\omega^{ab}$ one gets the
vielbein constraint, $T^a=De^a=0$. Of course, this is not a Yang-Mills action, since it is
not of a gauge invariant form (it is of the type $(dA+A\wedge A)\wedge A\wedge A$), and 
it is only invariant on-shell (if $T^a=De^a=0$), and if one identifies local translations
in the base space (diffeomorphisms with parameter $\lambda^{\mu}$) with local translations 
in the tangent space (gauge transformations with parameter $\lambda^a$) via the inverse
vielbein $e^{\mu}_a$, as $\lambda^{\mu}=e^{\mu}_a\lambda^a$. 

But in 3 dimensions, the Einstein action is gauge invariant, being a Chern-Simons theory,
\be
S_{EH}=\int_{M_3=\partial M_4} \epsilon_{abc}R^{ab}(\omega)\wedge e^c
=\int_{ M_4} \epsilon_{abc}R^{ab}(\omega)\wedge T^c
\ee
thus being of the type $\int_{M_4}F^A\wedge F^Bd_{AB}$, with $R_{ab}(\omega)$ and 
$T^c$ being the curvatures of the 3d Poincare group and $\epsilon_{abc}$ the symmetric group invariant $d_{AB}$. Witten noted this and used it to define the quantum theory 
of 3d gravity \cite{witten}. 

One can introduce a cosmological constant $\lambda/3$ by adding a term to the action
\be
S=S_{EH}+\frac{\lambda}{3} \int d^3 x \epsilon_{abc} e^a\wedge e^b\wedge e^c
\ee

The action is then invariant under the de Sitter (SO(3,1)) or Anti-de Sitter (SO(2,2))
group, depending on the sign of $\lambda$. In the following we will assume that $\lambda$
is positive, thus considering the AdS group. The invariance can be easily seen by rewriting 
the action in manifestly invariant form, 
\be
S=\int_{M_3=\partial M_4} (R^{ab}\wedge e^c+\frac{\lambda}{3}e^a\wedge e^b\wedge e^c) \epsilon_{abc}=\int_{M_4} \bar{R}^{ab}
\wedge T^c\epsilon_{abc}
\ee
where 
\be
\bar{R}^{ab}=R^{ab}(\omega)+\lambda e^a\wedge e^b
\ee
is the curvature of the SO(2,2) gauge field $A=\omega^{ab}J_{ab}+e^aP_a=\omega^{ab}J_{ab}
+\tilde{e}^a\tilde{P}_a$ and $e^a, P_a$ are quantities rescaled by $M=\sqrt{\lambda}$
($e^a=\tilde{e}^a/M, P_a=M\tilde{P}_a$), so that now for instance, $[P_a,P_b]=M^2 J_{ab}
=\lambda J_{ab}$.
The Wigner-Inonu contraction $M\rightarrow 0$ takes us to the previous case of the Poincare
group. 

In higher odd dimensions d=2n+1, one can define generalizations of the gravitational 
Chern-Simons action for the SO(2n,2) group (in the presence of $\lambda$), by 
\be
S=\int_{M_{2n+1}=\partial M_{2n+2}} I_{CS,2n+1}=\int_{M_{2n+2}} \bar{R}^{a_1a_2}\wedge ...
\wedge \bar{R}^{a_{2n-1}a_{2n}}\wedge T^{a_{2n+1}}\epsilon_{a_1...a_{2n+1}}
\ee
Here again the epsilon tensor $\epsilon_{a_1...a_{2n+1}}$ takes the role of the symmetric
tensor $d_{A_1..A_{n+1}}$ (=$Tr(T_{A_1}...T_{A_{n+1}})$ in the corresponding representation).
See also \cite{btz,tz1,tz2,zanelli,horava,nastase} for supergravity generalizations that
include this Chern-Simons gravity.

One can define a special gauge theory 
action (see the review \cite{zanelli} for a general discussion
of special, CS and BI, gravity actions)
defined by the epsilon tensor in even (d=2n) dimensions as well, by 
\be
S=\int_{M_{2n}} \bar{R}^{a_1a_2}\wedge ... \wedge \bar{R}^{a_{2n-1}a_{2n}}\epsilon
_{a_1...a_{2n}}
\ee
which is gauge invariant again (for the gauge group SO(2n-1,2)). This action is known 
as the Born-Infeld action, since one can formally understand it as the Pfaffian of the 
matrix $\bar{R}^{ab}=\bar{R}^{ab}_{\mu\nu}dx^{\mu}\wedge dx^{\nu}$, and since the square of 
the Pfaffian is the determinant, one can understand the above action formally as $\sqrt{\det
\bar{R}^{ab}}$, or Born-Infeld type. 

The CS and BI actions are particular cases of the so-called Lanczos-Lovelock actions for 
gravity (\cite{lanczos,lovelock}, see also \cite{zanelli} for a general discussion), 
of the type
\be
S_{LL}=\int \sum_{p=1}^{[D/2]}\alpha_p\epsilon_{a_1...a_D}R^{a_1a_2}\wedge...\wedge R^{a_{2p
-1}a_{2p}}\wedge e^{a_{2p+1}}\wedge ... \wedge e^{a_D}
\ee
The coefficients $\alpha_p '=\begin{pmatrix}&n\\&p\end{pmatrix}/(d-2p)$ in d=2n+1 and
$\alpha_p=\begin{pmatrix} &n\\&p\end{pmatrix}$ in d=2n define CS and BI actions, 
respectively. They are the unique cases in their dimension for which the LL action does not 
generate more constraints by acting with covariant derivatives on the equations of motion.

The uniqueness of the BI and CS actions make us hope that they can be obtained from each 
other via dimensional reduction. As noted in \cite{nastase}, 
it is pretty obvious how to do this  in 
the case of reducing CS to BI. 

One generalizes the dimensional reduction of the usual metric formulation of Einstein gravity
(and the corresponding supergravity method for the vielbein-spin connection formulation) 
to the case of CS gravity. The usual case of Kaluza-Klein (dimensional reduction of 5d 
gravity to 4d gravity + gauge field + scalar) has the metric reduction ansatz
\be
g_{\Lambda\Pi}=\phi^{-1/3}\begin{pmatrix} g_{\mu\nu}+\phi A_{\mu}A_{\nu}&\phi A_{\mu}\\
\phi A_{\nu}& \phi\end{pmatrix}
\ee
and as is well known one cannot put $\phi$ to 1 unless $A_{\mu}$ is also zero, otherwise 
the truncation is inconsistent, i.e. it doesn't satisfy the equations of motion. 
In terms of the vielbein $E_{\Lambda}^A$, that means that $E_5^A$ can be put to zero by a 
choice of gauge, $E_5^5=\phi^{1/3}$ is a scalar, $E_{\mu}^a=(E_5^5)^{-1/2}e_{\mu}^a$ 
with $e_{\mu}^a$
the 4d vielbein, and $E_{\mu}^5=A_{\mu}E_5^5$ completes the reduction ansatz. So if one puts
$E_5^5$ to 1 and $A_{\mu}$ to zero, that is a consistent trucation. In a first order 
formulation, with the spin connection independent of the vielbein, one would have to 
put also $\omega_{\mu}^{a5}$ and $\omega_5^{AB}$ to zero.

Here and in the following, when we dimensionally reduce we dentote by $\Lambda, \Pi, ...$
higher dimensional curved indices, by $\mu, \nu,...$ lower dimensional curved indices, by
$A,B,...$ higher dimensional flat (group) indices, and by $a,b,...$ lower dimensional flat
(group) indices. 

One then can generalize this KK reduction to the case of CS gravity, again choosing a gauge
so that $E_5^A=0$ and putting $E_5^5$ to 1 and the gauge field $E_{\mu}^a$ to zero, and see
what we get. It is easy to see that on the action, this procedure is equivalent to varying
the action with respect to $e_5^5$ (and then putting $\delta e_5^5$ to 1). 
As a result, the reduced Lagrangian is the same as the equation of motion for $e_5^5$
in the KK reduction background. As before,
one puts also $\omega_{\mu}^{a5}$ and $\omega_5^{AB}$ to zero. 

This KK reduction can be trivially be extended to arbitrary n, by putting $e_{2n+1}^{2n+1}=c$ 
and the rest of the extra dimensional fields to zero. Then one obtains
that $\alpha_p '$ becomes $\alpha_p$ by dimensional reduction, so that the CS action in 
d=2n+1 is dimensionally reduced to the BI action in d=2n. 

This procedure is a generalization of the usual KK reduction also because gravity is 
just a gauge field here, and the action is defined without the need of the inverse vielbein
or the star operation, thus is quasi-topological (it is genuinely topological in one 
dimension higher): it could actually be defined on an auxiliary space with a different metric!

More importantly however, now, even though we have put both the gauge field to zero and the  
scalar to one, we still have a truncation that is generically inconsistent. Specifically, 
we need to satisfy the equations of motion of the fields that were put to zero: consistency
of the truncation means that the lower dimensional equations of motion need to satisfy the 
higher dimensional equations of motion.

In the KK reduction background, we have $(\bar{R}^{a5})_{\mu\nu}=T^5=0$ (but $\bar{R}^{a5}
_{\mu 5}=\lambda c e^a_{\mu}\neq 0$), as well as $\bar{R}^{ab}_{\mu 5}
=T^a_{\mu 5}=0$. That can be easily seen to imply the $e_5^a$ and $\omega_5^{ab}$ equations 
of motion, as well as the $e^5_{\mu}$ and $\omega^{a5}_{\mu}$ equations of motion.
But the $e^5_5$ and $ \omega^{a5}_5$ equations of motion are not automatically satisfied,
and give instead (making an obvious generalization to d=2n+1 instead of d=5 that carries
through trivially)
\bea
&&\epsilon_{a_1...a_{2n}}\bar{R}^{a_1a_2}\wedge ... \wedge \bar{R}^{a_{2n-1}a_{2n}}=0\nonumber
\\&& \epsilon_{a_1...a_{2n}} \bar{R}^{a_1a_2}\wedge ... \wedge \bar{R}^{a_{2n-3}a_{2n-2}}\wedge
T^{a_{2n-1}}=0
\label{consist}
\eea
respectively. The first one says that the BI action in 2n dimensions is zero on-shell
(which was to be expected, since as mentioned the dimensional reduction of the action 
is equivalent to varying with respect to $e_d^d$, or the $e_d^d$ equation of motion), and 
the second can be seen to be satisfied by $T^a=0$, thus if we go to a second order formulation.
However, for consistency of the reduction, the question is whether the BI equation of motion,
\bea
&&\epsilon_{a_1...a_{2n}}\bar{R}^{a_1a_2}\wedge ... \wedge \bar{R}^{a_{2n-5}a_{2n-4}}\wedge
T^{a_{2n-3}}\wedge e^{a_{2n-2}}=0\nonumber\\
&& \epsilon_{a_1...a_{2n}}\bar{R}^{a_1a_2}\wedge ...\wedge \bar{R}^{a_{2n-3}a_{2n-2}}\wedge
e^{a_{2n-1}}=0
\label{bi}
\eea
satisfy the above conditions. For general d=2n, even $T^a=0$ is not the most general solution
of the $\omega_{\mu}^{ab}$ equation above, so in general neither of the conditions in 
(\ref{consist}) are satisfied and the reduction is generically inconsistent. 

Let us now go back to the case of interest, of 
n=2, thus 5d CS going to 4d BI gravity. In 4d, BI gravity
has the action
\be
\int d^4x \bar{R}^{ab}\wedge \bar{R}^{cd}\epsilon_{abcd}= \lambda\int d^4 x \epsilon_{abcd}
(\frac{1}{\lambda} R^{ab}\wedge R^{cd}+ 2R^{ab}\wedge e^c\wedge e^d +\lambda e^a\wedge e^b\wedge
e^c\wedge e^d)= S_{top}+S_{EH}+S_{\lambda}
\ee
thus is just the Einstein action, with a cosmological constant term, and a topological term
added (the Euler density), that will not affect the equations of motion! In fact, this exact 
topological term that takes Einstein gravity with $\Lambda$ into BI gravity, can be argued
to be needed for a variety of reasons. For instance, its addition regularizes Noether charges,
and defines a well posed boundary condition for locally AdS spacetimes, thus 
containing a built-in regularization \cite{acotz} (a similar situation is shown there to be  
true in higher dimensions, and other quantities are also fixed by the addition of the 
topological term). 

So at least at the classical level, the 5d CS action has a KK reduction background that reduces the theory to usual Einstein gravity! Unfortunately, as we saw, we need to satisfy the 
conditions (\ref{consist}) for consistency of the reduction. However, in 4d, $T^a=0$ (which 
solves the second equation in (\ref{consist})) is the most general solution of the $\omega_{\mu}
^{ab}$ BI equation of motion, so the second condition is satisfied now.
But to satisfy the first condition we need also to have the BI action 
equal to zero on-shell. Using the Einstein's equation (the $e_{\mu}^a$ BI equation
of motion), the first consistency condition (of zero BI Lagrangian) becomes 
\be
\epsilon_{abcd}(R^{ab}\wedge R^{cd}-\lambda^2 e^a\wedge e^b\wedge e^c\wedge e^d)=0
\label{4dconsist}
\ee
which integrated, would be $\lambda^2 \times volume =topological$ $ number$. 

So the KK reduction from 5d CS to 4d BI is still inconsistent, but now the only 
remaining consistency condition has a simple interpretation. Still, it is not
very nice to have a truncation be consistent only on a subset of the theory, so it would be 
very useful to find out if there is a consistent truncation. Whenever one has an inconsistent
truncation in dimensional reduction, there are 2 possible ways out: it may be possible to 
construct a nonlinear redefinition of fields that gives a consistent truncation (like for 
instance in the case of the $S_4$ and $S_7$ reductions of 11d supergravity to gauged supergravities in 7d and 4d, respectively \cite{dn,nvv}). 
Or one may need to keep more fields, like 
in the case of the original Kaluza-Klein reduction described before: keeping only $A_{\mu}$ 
and not $\phi$ is inconsistent, but by adding $\phi$ the KK reduction is consistent. Most 
known cases conform to either one of these situations, so it would be worthwhile exploring 
whether a nonlinear redefinition of fields, or allowing for more fields in the theory will 
make the KK reduction consistent, but we will leave it for further work. 

A hint that a nonlinear redefinition of fields making the truncation consistent is possible
comes from the work of Chamseddine \cite{chamseddine}
 (see also the discussion in \cite{nastase} for more details). 
Our ansatz has $e_5^5=c, e^a_5=\omega_5^{ab}=0, e^5_{\mu}=\omega^{a5}_{\mu}=0$, implying
$\bar{R}^{a5}_{\mu\nu}=T^5=0$, $\bar{R}_{\mu 5}^{ab}=T^a_{\mu 5}=0$, $\bar{R}^{a5}_{\mu 5}=
\lambda c e^a_{\mu}$ (and one needs (\ref{4dconsist})
for consistency of the reduction).
Chamseddine proved that the 5d CS gravity action has a classical background, satisfying 
$e_{0;5}^5=c, e_{0;5}^a=\omega_{0;5}^{ab}=0$, $ e_{0; \mu}^5=\omega_{0;\mu}^{a5}=0$, that 
implies
$\bar{R}^{a5}_{0;\mu\nu}=T_0^5=0$, $\bar{R}^{ab}_{0;\mu 5}=T^a_{0;\mu 5}=0$, 
$\bar{R}^{a5}_{0;\mu 5}
=\lambda c e^a_{0;\mu}$, but also $T_0^a=0, \bar{R}^{ab}_0=0$ (thus the background satisfies 
our condition (\ref{4dconsist})
 for consistency of the reduction). Around this classical background, fluctuations
have a quadratic action that is exactly that of the 4d Einstein action with cosmological term,
in either the first or the second order formulations. Since as we mentioned, the topological 
Euler density doesn't contribute to the equations of motion, this is exactly what we expect 
from a consistent truncation extension of our ansatz. Moreover, in the Chamseddine background, 
the fluctuations for the extra 5d fields drop out, so in effect we have 
$e_5^5=c, e^a_5=\omega_5^{ab}=0, e^5_{\mu}=\omega^{a5}_{\mu}=0$ for the full fields, not 
only for the background, exactly as in our case. In conclusion, this makes it likely
that a consistent truncation extension of our ansatz can be found.

\section{New dimensional reduction: from D=2n+1 CS to D=2n-1 CS gravity via "descent 
equations"}

As we saw in the previous section, the CS gravity action in d=2n+1 reduces to the BI gravity 
action in d=2n
by putting $e_{2n+1}^{2n+1}=c$ and all the rest of the extra dimensional fields to zero. 
At the level of the Lanczos-Lovelock-type action, CS gravity looks schematically like
\be
{\cal L}=\sum_p \alpha '_{p;n} (R)^{\wedge p} \wedge (e)^{\wedge (2n+1-2p)};\;\;\;
\alpha _{p;n}' = \frac{1}{2n+1-2p}\frac{n!}{p! (n-p)!}
\ee
and since as mentioned, the dimensional reduction is equivalent to varying with respect to
$e_{2n+1}^{2n+1}$ (its equation of motion), the result is
\be
{\cal L}=\sum_p \alpha _{p;n} (R)^{\wedge p} \wedge (e)^{\wedge (2n-2p)};\;\;\;
\alpha _{p;n} =\alpha_{p;n}'(2n+1-2p)=\frac{n!}{p! (n-p)!}
\ee 
which is BI gravity in d=2n. However, if we continue to dimensionally reduce in the same
way, by putting $e_{2n}^{2n}=c$ and the rest of the extra dimensional fields to zero, we
do not get the CS theory in d=2n-1, but rather
\be
\tilde{\alpha}_{p;n-1}= (2n-2p)\alpha_{p;n}=2n\alpha_{p;n-1} 
\neq \alpha _{p;n-1}'
\ee
thus a different Lanczos-Lovelock Lagrangian in d=2n-1, which is 
specifically,
\be
\tilde{S}
=2n\int d^{2n-1}x \epsilon_{a_1...a_{2n-1}}\bar{R}^{a_1a_2}\wedge ...\wedge \bar{R}^{a_{2n-3}
a_{2n-2}}\wedge e^{a_{2n-1}}
\label{newaction}
\ee

So we cannot obtain a sequence of dimensional reductions $CS\rightarrow BI\rightarrow CS$
as one might have suspected, but one may still ask whether one can still dimensionally 
reduce the CS action in d=2n+1 to a  CS action in d=2n-1. Luckily, for that reduction there
exist an analog for the dimensional reduction of usual CS gauge theories, the "descent 
equations". 

The descent equations are usually written formally
as $\omega_{D+2}=d\omega_{D+1}$, $\delta
\omega_{D+1}=d\omega_D$, where $\omega_D$ is an anomaly term in even dimensions, that
integrated gives an action. This means that one "descends" (dimensionally reduces) from 
an anomaly term to another. 
Concretely, for a single gauge field A with field strength $F=dA, 
dA=0$, one can consider the form $\omega_{2n+2}=Tr F^{\wedge (n+1)}$. Because $d\omega_{2n+2}
=0$, one can write locally $\omega_{2n+2}=d\omega_{2n+1}$, with $\omega_{2n+1}$ the 
Chern-Simons form. Then, because $\delta \omega_{2n+2}=(n+1)d tr(\delta A F^n)$, it 
means that $\delta \omega_{2n+1}=tr(\delta A F^n)$. Then, under a gauge transformation
$\delta A=D\Lambda$, one obtains $\delta_{gauge,\Lambda}\omega_{2n+1}= d tr(\Lambda F^{n})
$, and upon eliminating $\Lambda$ we get $d\omega_{2n}$, which is as we mentioned what one 
usually understands by descent equations. 

But we will instead understand it (for our purposes) 
as $\omega_{2n+2}=d\omega_{2n+1}, \delta \omega_{2n+1}
=tr(\delta A F^n)$, which equals $\omega_{2n}$ if we put $\delta A=1$. We can notice already 
the similarity to what we did before, when we dimensionally reduced by putting $e_D^D$ to 1 and 
said that this is the 
same as varying with respect to $e$ and putting the variation to 1. 

However, this is not the whole story, as this formalism doesn't apply automatically . We 
note that for dimensional reduction of gravity, we must also "dimensionally reduce" the
gauge group, in this case the AdS group SO(d-1,2), and in the usual descent equations formalism
the gauge group is maintained, so the analysis will not automatically carry over, we have
to define exactly at which step and how we reduce the gauge group. Moreover, in our case, we 
have a choice of what gauge field to use for $\delta A=1$, either $e_{\mu}^a$ or
$\omega_{\mu}^{ab}$. In view of the previous section, we might think using the vielbein is 
required, but in fact we will see that is not the case. 

So starting with the CS form (action) in d=2n+1, we want to put a gauge field to 1, equivalent
to varying A and putting $\delta A=1$. Specifically, we put $\omega_{2n+1}^{2n,2n+1}=c$ and 
the rest of the extra-dimensional fields to zero, i.e. $\omega_{\Lambda}^{a, 2n}=\omega_{\Lambda
}^{a,2n+1}=0$, $e_{\Lambda}^{2n}=e_{\Lambda}^{2n+1}=0$, $\omega_{\mu}^{2n, 2n+1}=e_{2n+1}^a
=\omega_{2n+1}^{ab}=0$. It is then easy to check (as it should be obvious from the descent
equations formalism above) that then the d=2n+1 CS gravity action reduces in d=2n to 
\bea
&&S_{2n}(M_{2n})=\int_{M_{2n}}
 d^{2n}x\epsilon_{a_1....a_{2n-1}}\bar{R}^{a_1a_2}\wedge ... \wedge 
\bar{R}^{a_{2n-3}a_{2n-2}}\wedge T^{a_{2n-1}}\nonumber\\&&
= \int_{M_{2n}}d^{2n}x dI_{CS, 2n-1}=\int_{M_{2n-1}=\partial 
M_{2n}} d^{2n-1}x I_{CS,2n+1} =S_{CS,2n-1}(M_{2n-1})
\eea
thus a further dimensional reduction is obtained by just going to the 2n-1 dimensional 
boundary of the 2n dimensional space and obtaining the 2n-1 dimensional CS action. 

Again, we need to check the consistency of the dimensional reduction from d=2n+1 to d=2n. 
We have $\bar{R}^{2n,2n+1}=\bar{R}^{a, 2n}=\bar{R}^{a,2n+1}=T^{2n}=T^{2n+1}=0$, $\bar{R} 
^{ab}_{\mu, 2n}=T^a_{\mu. 2n}=0$. Using this, it 
is easy to check that the $\omega_{\Lambda}^{a,2n}, \omega_{\Lambda}^{a,2n+1}, e_{\Lambda}^{2n},
e_{\Lambda}^{2n+1}$, and $\omega_{\mu}^{2n,2n+1}, e_{2n+1}^a,\omega_{2n+1}^{ab}$ equations 
of motion are satisfied. The only nontrivial equation of motion is, as for the CS to BI 
reduction, the equation of motion for the nonzero field, $\omega_{2n+1}^{2n,2n+1}$, which 
gives
\be
\epsilon_{a_1...a_{2n-1}}\bar{R}^{a_1a_2}\wedge ...\wedge \bar{R}^{a_{2n-3}a_{2n-2}}\wedge
 T^{a_{2n-1}}=0
 \ee
 or that the reduced Lagrangian is zero on-shell. As the lower dimensional action (in d=2n)
 is topological, there are no equations of motion to help us solve the consistency 
 condition. 
 
 So the KK reduction is again inconsistent in general, but now it becomes consistent if 
 we just go to a first order formulation (for $T^a=0$). Still, it would also be nice to find 
 whether there exists a way to make the reduction consistent, either by making a nonlinear
 field redefinition, or by keeping more fields.
 
Let us consider what would be a possible condition for finding a consistent reduction. 
We will search for a reduction directly from 2n+1 to 2n-1 dimensions.
An automatically consistent reduction would be found if one reduces the equations of motion 
directly. The CS equations of motion in d=2n+1 are 
\bea
&& \epsilon_{a_1...a_{2n+1}}\bar{R}^{a_1a_2}\wedge ...\wedge \bar{R}^{a_{2n-1}a_{2n}}=0
\nonumber\\&&
\epsilon_{a_1...a_{2n+1}}\bar{R}^{a_1a_2}\wedge ... \wedge \bar{R}^{a_{2n-3}a_{2n-2}}\wedge
T^{a_{2n-1}}=0
\label{cs}
\eea
Then the d=2n+1 CS equations of motion reduce to the d=2n-1 CS equations of motion if 
\bea
&&\bar{R}^{d-1,d}_{d-1,d}=c;\;\; \bar{R}_{\mu ,d-1}^{ab}=\bar{R}_{\mu ,d}^{ab}=\bar{R}^{ab}_{
d,d-1}=0;\;\; 
\bar{R}^{d,d-1}_{\mu\nu}=\bar{R}^{d,d-1}_{\mu d}=\bar{R}^{d,d-1}_{\mu 
d-1}=0\nonumber \\
&&T^a_{\mu, d-1}=T^a_{\mu,d}=T^a_{d,d-1}=0;\;\;\;\bar{R}^{a,d-1}=\bar{R}^{a,d}=T^{d-1}=T^{d}=0
\label{cscs}
\eea
as well as $\bar{R}^{ab}_{\mu\nu}=\bar{R}^{ab}_{\mu\nu, red}, T^a_{\mu\nu}=T^a_{\mu\nu,red}$.
However, satisfying these conditions, which are 
written in gauge invariant way, in terms of the curvatures
(field strengths) of the AdS group SO(d-1,2), by an explicit choice of gauge fields 
$e_{\Lambda}^A$ and $\omega_{\Lambda}^{AB}$ is quite difficult, as we will see in the next 
section. We will find the same conditions for the reduction from d=2n+2 to d=2n BI gravity.

\section{D=2n+2 BI to D=2n BI?}

We have established that CS gravity reduces to BI gravity, and CS gravity in d=2n+1 reduces 
to CS in d=2n-1, albeit the issue of consistency of 
the reduction is not settled, and that we cannot reduce in a simple way BI to CS gravity. 
But we want to analyze in more detail the BI to CS reduction, as well as the possibility of 
reduction of BI in d=2n+2 to BI in d=2n. 

In order to find a consistent reduction of the BI theory in d=2n+2 to the CS theory in 
d=2n+1, we need to dimensionally reduce the corresponding equations of motion. We 
see that an embedding of the CS equations of motion (\ref{cs}) into the BI equations 
of motion (\ref{bi}) 
(for $n\rightarrow n+1$) is obtained if we can satisfy the following conditions on the 
SO(d-1,2) group curvatures (field strengths)
\be
\bar{R}^{a,2n+2}=R^{a,2n+2}+\lambda e^a\wedge e^{2n+2}=0; \;\; T^{2n+2}=0; \;\;
\bar{R}^{ab}_{\mu , 2n+2}=\bar{T}^a_{\mu, 2n+2}=0
\ee
together with the condition that $e^{2n+2}$ is nonzero
and depends only on $x^{2n+2}$ and the fact that 
$R^{ab}_{\mu\nu}=R^{ab}_{\mu\nu, red}$ and $T^a_{\mu\nu}=T^a_{\mu\nu, red}$.

We try to satisfy them by using (the gauge condition) $e_{2n+2}^a=0$ and putting the gauge 
field $e^{2n+2}_{\mu}$ to zero also (that solves $T^a_{\mu\nu}=T^a_{\mu\nu, red}$),
 thus having only $e_{2n+2}^{2n+2}, \omega_{\mu}^{a,2n+2},
\omega_{2n+2}^{ab}$, $\omega_{2n+2}^{a,2n+2}$ nonzero. 

Then from $R^{ab}_{\mu\nu}=R^{ab}_{\mu\nu, red}$ we get $\omega_{[\mu}^{a,2n+2}\omega_{\nu]}
^{2n+2,b}=0$ which is solved by $\omega_{\mu}^{a,2n+2}=0$, that also solves $\bar{R}^{a, 2n+2}
_{\mu\nu}=0$ and $T^{2n+2}_{\mu\nu}=0$. However, then $T^{2n+2}_{\mu, 
2n+2}=0$ gives $\omega_{2n+2}^{2n+2,a}e_{\mu}^a=0$, so $\omega_{2n+2}^{2n+2,a}=0$, which
contradicts the condition that $\bar{R}^{a,2n+2}_{\mu, 2n+2}=0$, since $R^{a,2n+2}_{\mu, 2n+2}
=0$, but $e^{2n+2}_{2n+2}e_{\mu}^a$ is nonzero.

So it seems there is no solution of this type. One could try to relax the condition that 
$R^{ab}_{\mu\nu}=R^{ab}_{\mu\nu, red}$ and try to find another way to embed the CS equations
in the BI equations, but I have not been able to find one, and it seems unlikely to be 
possible. The simplest possibility is that we allow for a redefinition of $\lambda$, 
by having $R^{ab}_{\mu\nu}=R^{ab}_{\mu\nu,red}+k e^a_{[\mu}e^b_{\nu]}$, which when 
substituted in $\bar{R}^{ab}_{\mu\nu}$ gives a rescaling $\lambda\rightarrow \lambda+k$,
but this does not work either.

However, one can instead keep also $e_{2n+2}^a$ and $e^{2n+2}_{\mu}$
(even though this was not needed in usual KK reduction), but then the equations
become too complicated, and I did not find a solution. It is also possible that global 
issues will allow nontrivial gauge fields $e^A_{\Lambda}, \omega^{AB}_{\Lambda}$, 
defined on patches.

Let us turn to the possibility of finding a consistent reduction of the BI equations of 
motion in d=2n+2 to the ones in d=2n (\ref{bi}). Similarly, the conditions on the 
SO(d-1,2) group curvatures are now
\bea
&&\bar{R}^{2n+1,2n+2}_{2n+1,2n+2}=c;\;\; \bar{R}_{\mu ,2n+1}^{ab}=
\bar{R}_{\mu ,2n+2}^{ab}=\bar{R}^{ab}_{
2n+1,2n+2}=0;\nonumber\\
&& \bar{R}^{2n+1,2n+2}_{\mu\nu}=\bar{R}^{2n+1,2n+2}_{\mu, 2n+2}=
\bar{R}^{2n+1,2n+2}_{\mu, 2n+1}=0
\nonumber \\
&&T^a_{\mu, 2n+1}=T^a_{\mu,2n+2}=T^a_{2n+1,2n+2}=0;\;\;\;\bar{R}^{a,2n+1}=\bar{R}^{a,2n+2}=T^{2n+1}=T^{
2n+2}=0
\label{bibi}
\eea
which is the same as the condition for going from d=2n+1 CS to d=2n-1 CS (\ref{cscs}), 
taking into account the change in dimension. We will also add $T^a_{\mu\nu}=T^a_{\mu\nu, red}$,
but we relax the condition that $R^{ab}_{\mu\nu}=R^{ab}_{\mu\nu, red}$ as above, by allowing 
for a rescaling of $\lambda$ via $R^{ab}_{\mu\nu}=R^{ab}_{\mu\nu,red}+ke^a_{[\mu}e^b_{\nu ]}$.

We look for solutions with nonzero $e_{2n+1}^{2n+1}, e^{2n+2}_{2n+2},e^{2n+1}_{2n+2},
e^{2n+2}_{2n+1}$ and $\omega_{2n+1}^{2n+1,2n+2},\omega_{2n+2}^{2n+1,2n+2}$, $\omega_{\mu}
^{a,2n+1},\omega_{\mu}^{a,2n+2}$, and all of them being only functions of $x^{\mu}$ (dimensionally reduced coordinates) only. The last two ($\omega_{\mu}^{a, 2n+1}, \omega_{\mu}^{
a, 2n+2}$) need to be nonzero in order to find a nonzero $R^{a,2n+1}_{\mu, 2n+1}, R^{a,2n+2}
_{\mu, 2n+2}$ as needed. Then the condition that 
$R^{ab}_{\mu\nu}=R^{ab}_{\mu\nu,red}+ke^a_{[\mu}e^b_{\nu ]}$ gives $\omega^{a,2n+1}_{\mu}=
\beta e^a_{\mu}, \omega^{a, 2n+2}_{\mu}=\gamma e^a_{\mu}$, which in turn implies that 
$R^{a,2n+1}_{\mu\nu}=\beta T^a_{\mu\nu}, R^{a, 2n+2}_{\mu\nu}=\gamma T^a_{\mu\nu}$ thus 
are zero on-shell (for $T^a=0$), as needed. Then the conditions (\ref{bibi}) become
(assuming that $e_m^i,\; i,m=2n+1,2n+2$ are constants)
\bea
&&\omega^{2n+2,2n+1}_{2n+1}=-\frac{\lambda}{\gamma}e_{2n+1}^{2n+1};\;\;
\omega^{2n+1,2n+2}_{2n+2}=-\frac{\lambda}{\beta}e^{2n+2}_{2n+2};\nonumber\\
&& e^{2n+1}_{2n+2}=-\frac{\gamma}{\beta}e^{2n+2}_{2n+2};\;\;\; e^{2n+2}_{2n+1}=-\frac{\beta}{
\gamma}e^{2n+1}_{2n+1}
\eea
which solves everything, except one condition: now $e^{2n+1}_{[2n+1}e^{2n+2}_{2n+2]}=0$,
thus we obtain that $\bar{R}^{2n+1,2n+2}_{2n+1,2n+2}=0$ instead of a nonzero constant.

Thus the natural simplest guess doesn't work. However, now there are many more generalizations
to be tried: we can make the above fields depend also on $x^{2n+1},x^{2n+2}$, and we can 
reintroduce more fields: $e_{\mu}^{i}, e^a_{m}$, $\omega^{a,i}_m, \omega^{2n+1,2n+2}_{\mu}$, 
with $i,m=2n+1,2n+2$. Unfortunately, then the equations required become prohibitively 
difficult to solve. 

A natural question to ask is whether we can also use a dimensional reduction based on a 
version of descent equations for the BI gravity, like we did for CS gravity. The two would not 
be the same a priori, since as mentioned the descent equations need to be supplanted by a 
prescription about how to dimensionally reduce the gauge group also, and moreover the 
d=2n Lagrangian in the CS case is not the BI Lagrangian. 

The descent equations would be formally of the type $I_{2n+2}=dI_{2n+1}, \delta I_{2n+1}/\delta
A=I_{2n}$, and as we saw in the last section, the gauge field that is put to one (corresponding
to $\delta A$) needs to be $\omega_{2n+1}^{2n+1,2n+2}$ (it was actually $\omega_{2n+1}^{2n,
2n+1}$ in the last section because the gauge group was different in d=2n). It will actually
turn out that the first step is the hardest (writing $I_{2n+2}$ as an exact form), so we will
instead start with the second step. The BI Lagrangian in d=2n is schematically of the 
type (ignoring indices)
\be
{\cal L}_{2n}=\sum_{p=0}^n\frac{n!}{p!(n-p)!}\epsilon 
R^{\wedge p}\wedge (e\wedge e)^{\wedge (n-p)}
\ee
so by integrating with an $\delta \omega$ it is easy to check (since $\int \delta\omega
R(\omega)^{\wedge p}\sim S_{2p+1, CS}(\omega)$ and one can explicitly check a few 
examples) that the d=2n+1 Lagrangian that 
gives the d=2n one by $\omega_{2n+1}^{2n+1,2n+2}=1$ is
\bea
{\cal L}_{2n+1}&=& \sum_{p=0}^n\frac{n!}{(p+1)p!(n-p)!}\epsilon\; "I_{2p+1,CS}(\omega)"\wedge 
(e\wedge e)^{\wedge(n-p)}\nonumber\\
&=& \sum_{p=0}^n\frac{n!}{(p+1)!(n-p)!}\epsilon_{a_1...a_{2n+2}}I_{2p+1,CS}^{a_1...a_{2p+2}
}(\omega)\wedge e^{a_{2p+3}}\wedge ...\wedge e^{a_{2n+2}}
\label{newcs}
\eea
where in the first line we wrote the action schematically and in the second line 
$I_{2p+1,CS}^{a_1...a_{2p+2}}
(\omega)$ is the integrand that contracted with $\epsilon_{a_1...a_{2p+2}}$ (for gauge group
SO(2p+2)) would give the corresponding CS form. Of course, on the first line, we don't have
a CS form, since although it is contracted with an epsilon, the sum runs over 2n+2 indices
instead of 2p+2. If we would actually have a CS form, then we would have schematically
\be
d{\cal L}_{2n+1}=\sum_{p=0}^n\frac{n!}{(p+1)!(n-p)!}\epsilon R^{\wedge (p+1)}(\omega)\wedge
(e\wedge e)^{\wedge (n-p)}+{\rm terms \; with} \; de
\label{dnewcs}
\ee
to be compared with the BI form minus the cosmological constant term
\be
{\cal L}
_{2n+2}-e^{\wedge (2n+2)}=(n+1)\sum_{p=0}^n\frac{n!}{(p+1)!(n-p)!}R^{\wedge (p+1)}(\omega)
\wedge(e\wedge e)^{\wedge (n-p)}
\label{higherbi}
\ee
thus we would obtain all terms in ${\cal L}
_{2n+2}$ except the cosmological constant term, and we 
would get extra $de$ terms. As mentioned however, we don't have the actual CS form, 
and consequently we also don't have the $I_{2p+2}(\omega)$ form in $d{\cal L}_{2n+1}$, 
for the same reason,
that in the epsilon group contraction, the sum runs over 2n+2 instead of 2p+2 indices.
As a result for instance, the terms of type $\epsilon_{a_1...a_{2n+2}} (\omega\wedge\omega)
^{a_1a_2}\wedge ...\wedge (\omega\wedge \omega)^{a_{2p+1}a_{2p+2}}\wedge e^{a_{2p+3}}\wedge
...\wedge e^{a_{2n+2}}$ are not zero, and they cannot be obtained from a derivative of 
something. In a $I_{2p+2}(\omega)$ form, the terms 
$\epsilon_{a_1...a_{2p+2}} (\omega\wedge\omega)
^{a_1a_2}\wedge ...\wedge (\omega\wedge \omega)^{a_{2p+1}a_{2p+2}}$ are actually zero by 
symmetry, as can be easily seen using a general formalism. In $I_{2n}=F^{A_1}\wedge...\wedge
F^{A_n}t_{A_1...A_n}$, with $t_{A_1...A_n}=tr(T_{A_1}...T_{A_n})$ in the corresponding 
representation, the term with no $dA$'s can be rewritten as $A^{B_1}\wedge A^{C_1}\wedge...
\wedge A^{B_n}\wedge A^{C_n}tr(T_{B_1}T_{C_1}...T_{B_n}T_{C_n})$, and while the trace is 
cyclically symmetric, the gauge fields multiplying it are antisymmetric. 

A last hope to get the right result would be to rewrite the terms with $de$ by using the
torsion constraint, $T=de+\omega\wedge e=0$ (thus on-shell),
which would generate a lot of terms with 
extra $\omega$'s, including ones with no derivatives at all. But first of all there will be 
too many terms then, and even in the lowest dimensional relevant case (when the type of 
terms matches), for d=4 reduced to d=2, we don't get the correct result. Then 
we have
\bea
&&{\cal L}_3=\epsilon_{abcd}\omega^{ab}\wedge (d\omega^{cd}+\frac{2}{3}\omega^{ce}\wedge 
\omega^{ed}+2e^c\wedge e^d)\nonumber\\
&&d{\cal L}_3=\epsilon_{abcd}(d\omega^{ab}\wedge d\omega^{cd}+2\omega^{ab}\wedge(\omega^{ce}
\wedge \omega^{ed})\nonumber\\&&
+2d\omega^{ab}\wedge e^c\wedge e^d+4\omega^{ae}\wedge \omega^{eb}\wedge
e^c\wedge e^d) 
\eea
whereas ${\cal L}_4$ has a 2 instead of a 4 multiplying  the last term and of course the
extra $\epsilon_{abcd}e^a\wedge e^b\wedge e^c\wedge e^d$ term. To get the second line we
have used $T^a=0$ to replace $de^a$ with $-\omega^{ab}\wedge e^b$, and used antisymmetry
identities to recouple the indices.

Let us mention that although (\ref{dnewcs}) doesn't match (\ref{higherbi}), a lot of it
does: the term with p=n, with no vielbein and only spin connections obviously works, since
it is just due to a usual type of descent equation. Also the terms with only $d \omega$'s 
and no $\omega$'s or $de$'s work (as should be obvious from the $I_{2p+1,CS}(\omega)$ 
analogy in (\ref{newcs})). It is also clear that the cosmological constant term 
$\epsilon_{a_1... a_{2n+2}}e^{a_1}\wedge ...\wedge e^{a_{2n+2}}$ cannot be obtained, since
it has no derivatives or $\omega$'s. In the usual descent equations, the term with only 
A's and no dA's
is zero by symmetry as we saw, but now we cannot interpret $\epsilon_{a_1...a_{2n+2}}$
as $tr(T_{B_1}T_{C_1}...T_{B_{n+1}}T_{C_{n+1}})$, since the epsilon term is cyclically 
antisymmetric, whereas the trace is cyclically symmetric (which is the origin of the 
vanishing of the this term in the usual descent equations). 

In conclusion, we see that the descent equation formalism 
for reducing BI to BI doesn't work, and the consistent
reduction of the equations of motion has no simple solution, although a complicated one 
might exist.

\section{Discussion and conclusions}

In this paper I have analyzed the possible dimensional reductions between the Chern-Simons
(in d=2n+1) and Born-Infeld (in d=2n)
gravity theories. These are gauge theories of the AdS groups SO(d-1,2), with 
gauge fields the vielbein and spin connection. The fact that they are defined as gauge theories
means that in principle one can think of them as being defined in auxiliary spaces, and the 
vielbein and spin connection treated as regular gauge fields on the space. 
The important fact is that the 4d Born-Infeld theory is just the Einstein theory in 
first order formulation for the vielbein and spin connection, with a cosmological constant term
and an extra topological term (the 4d Euler density). Thus classically, this theory is the 
same as Einstein theory. The addition of the particular topological term is needed for a 
variety of reasons, see for instance \cite{acotz}.

I have found that the 4d BI theory can be obtained by a generalized dimensional reduction 
of the 5d CS theory (a reduction modelled after the usual KK reduction in first order 
formalism), which has $e_5^5=c$ (constant) and the rest of the extra fields to zero. 
The reduction is only consistent if the local version of a global condition,
eq. (\ref{4dconsist}), is satisfied. A result by Chamseddine \cite{chamseddine}
 showing that the fluctuations
around a similar background of 5d CS theory match fluctuations around 4d Einstein theory, 
suggest that there should be a way to make the reduction consistent (either by a nonlinear 
redefinition of fields as in \cite{dn,nvv}
 or by the addition of more fields as for the original Kaluza
Klein reduction). The reduction generalizes to CS theory in d=2n+1 reduced to BI theory in 
d=2n, but the consistency conditions are more complicated.

I have shown that the reverse reduction, of d=2n+2 BI theory, using the same ansatz,
does not result in the d=2n+1 CS theory, but rather in a new action (\ref{newaction}). 
Instead, I have shown that one can use a new type of dimensional reduction,  which is a 
generalization of the descent equations formalism, $"\delta I_{2n+1}/\delta A=I_{2n}",
I_{2n}=dI_{2n-1}$. Specifically, the first step is done by putting $\omega_{2n+1}^{2n,2n+1}=1$
and the rest of the extra fields to zero, which does not have an analog in usual dimensional
reduction. The second step is just reducing a the d=2n topological theory on the boundary 
to the d=2n+1 CS theory. This is a generalization of the descent equations formalism, since one
needs to define a "dimensional reduction of the gauge group" as well (in the usual descent
equations, the gauge group stays the same). The dimensional reduction is found to be 
consistent if the reduced Lagrangian is zero on-shell, or if we go to a second order formulation
by $T^a=0$. I also gave an ansatz for the SO(d-1,2) field strengths (curvatures)
 that would automatically 
 give a consistent reduction of d=2n+1 to d=2n-1 CS gravity, if one could find a 
 set of gauge fields that satisfy these conditions. 

For the reduction from d=2n+2 BI to d=2n+1 CS gravity, or directly from d=2n+2 BI to d=2n BI
gravity, I have given also ansatze for the field strengths, that if satisfied by a set of 
gauge fields would automatically give a consistent reduction. Unfortunately, I was not able 
to find solutions to them: the natural simple ansatze that I have tried do not work. 
I have also tried a generalization of the descent equation formalism for the reduction from
d=2n+2 BI to d=2n BI gravity (which is not a priori the same as the one for d=2n+1 CS to 
d=2n-1 CS, as there the d=2n action is not BI, and also one needs to define the "dimensional
reduction of the gauge group"). I have found an action in d=2n+1, (\ref{newcs}) that can be 
reduced to the BI action in d=2n via $\omega_{2n+1}^{2n+1,2n+2}=c$, however the derivative of
that integrand does not give the d=2n+2 BI integrand, but rather comes close. 

Let us comment now on the possible implications of this work. As mentioned, the BI and CS
actions are gauge theories of the AdS groups SO(d-1,2), with generators $P^a$ and $J^{ab}$.
When dimensionally reducing d=2n+1 CS theory to d=2n-1 CS theory, the intermediate d=2n 
theory we considered is a topological theory. Thus the 4d BI theory, which is the Einstein 
theory in the presence of a cosmological constant, and with a topological term (the 4d Euler
density) added, is embedded in higher dimensional topological theories, via the CS theories.
This is similar to the 3d case analyzed by Witten \cite{witten}, 
where Einstein gravity actually is of 
CS type (and can thus be embedded in a 4d topological theory), and this helped in defining 
a good and renormalizable quantum gravity theory. The hope is that one can use the same 
techniques now to help define 4d Einstein quantum gravity, although of course how that can be 
done remains an open question. 

Note that by successive dimensional reductions the 4d BI gravity theory can be 
be embedded in 11d CS gravity theory with SO(10,2) gauge group. 
In \cite{nastase,horava} it was argued that M theory should have an
invariance group $OSp(1|32)\times OSp(1|32)$
(the necessity of $OSp(1|32)$ as a possible invariance group for 11d supergravity 
was already argued for in the original paper
\cite{cjs}), and a Chern-Simons supergravity for this
group was constructed, that included the SO(10,2) Chern-Simons gravity as the purely 
gravitational part (see also \cite{btz,tz1,tz2}).  
And because experimentally the cosmological constant
is small, and $\lambda $ measures the size of terms with many R's with respect to terms with
fewer R's, physically we are in the $\lambda\rightarrow 0$ limit, or high energy limit, 
in which it was argued in \cite{nastase}
 that usual 11d supergravity should be obtained. In particular, 
in this limit, the $OSp(1|32)\times OSp(1|32)$ group contracts to the group of d'Auria and
Fre \cite{df}, describing usual 11d supergravity (almost) as a gauge theory. 

Finally, it should be noted that the d=2n topological actions with gauge group SO(2n-2,2)
that reduce on the boundary to d=2n-1 CS have a natural interpretation as a pure spin
connection theory in a space with 
signature (2n-2,2). Indeed, with $\Omega^{ab}=\omega^{ab}, \Omega^{a,2n}=e^a, a=1,2n-1$, the 
curvatures $R^{AB}(\Omega)=d\Omega^{AB}+\Omega^{AC}\wedge \Omega^{CB}$ split into 
$R^{ab}(\Omega)=\bar{R}^{ab}(\omega)$ and $R^{a,2n}(\Omega) =T^a$, and now $\Omega$ can \
be actually interpreted as a spin connection on the base space. In the paper we have used
$\omega^{ab}$ and $e^a$ (where a=1,2n-1) for this d=2n topological action 
as just SO(2n-2,2) gauge fields on the base space with signature (2n-1,1), but then they 
cannot have the interpretation of spin connection and vielbein on the base space! In particular,
the index $a$ runs over 2n-1 values instead of 2n.
Note that for the BI and CS theories, one can however interpret $\omega^{ab}$ and $e^a$ as 
spin connection and vielbein, respectively (the indices run over the correct number of 
values).

The use of (2n-2,2) signature, and the interpretation of the theory as pure spin connection 
theory might appear unusual for general n. But for n=6, i.e. (10,2) dimensions, there is a
reason for taking it under consideration. The maximal supergravity in 4d is ${\cal N}=8$, and 
the 8 gravitini become a single one in both (10,1) and (10,2) dimensions, i.e. the  theory 
with minimal supersymmetry in both (10,1) and (10,2) dimensions would have the maximal 
${\cal N}=8$ in 4d. Therefore many people have searched for a supergravity theory in (10,2)
dimensions (see for instance \cite{cfgpv2}),
 but of course the presence of two times with its known acausality issues 
makes that search problematic for an usual theory of gravity. But now the theory 
in (10,2) dimensions would have only a spin connection and no vielbein, and moreover it 
would be topological! So it certainly makes sense, and perhaps one can find also a 
supersymmetric version (perhaps as in \cite{nastase}) 
that would be related to usual 11d supergravity 
or ${\cal N}=8$ 4d supergravity.

{\bf Acknowledgements} 
This research has been done with support from MEXT's program
"Promotion of Environmental Improvement for Independence of Young Researchers"
under the Special Coordination Funds for Promoting Science and Technology.


\newpage

\begin{thebibliography}{99}

\bibitem{bdr} Z.Bern, L.J. Dixon and R. Roiban, "Is N=8 supergravity ultraviolet finite?"
Phys. Lett. B644 (2007) 265 and hep-th/0611086;
Z. Bern, J.J. Carrasco, L.J. Dixon, H. Johansson, D.A. Kosower, R. Roiban,
"Three-loop superfiniteness of N=8 supergravity," hep-th/0702112

\bibitem{grv} M.B. Green, J.G. Russo and P. Vanhove, "Non-renormalisation conditions 
in type II string theory and maximal supergravity," hep-th/0610299 and
"Ultraviolet properties of maximal supergravity," hep-th/0611273 

\bibitem{peter} P. van Nieuwenhuizen, "Supergravity," Phys. Rept. 68 (1981) 189

\bibitem{witten} E. Witten, ``2+1 dimensional gravity as an exactly 
soluble system'', Nucl.Phys. B 311 (1988) 46

\bibitem{pvn} P. van Nieuwenhuizen, ``Three-dimensional conformal 
supergravity and Chern-Simons terms'', Phys. Rev. D 32 (1985) 872

\bibitem{at} A. Achucarro and P.K. Townsend, ``A Chern-Simons action 
for three-dimensional Anti-de Sitter supergravity theories'', Phys. Lett. B 
180(1986) 89

\bibitem{dm} A. Dimakis and F. Mueller-Hoissen, "Clifform calculus with applications 
to classical field theories," Class. Quant. Grav. 8 (1991) 2093.

\bibitem{mh} F. Mueller-Hoissen, "From Chern-Simons To Gauss-Bonnet,"
Nucl. Phys. B346 (1990) 235.

\bibitem{nastase} H. Nastase, "Towards a Chern-Simons M-theory of $OSp(1|32)\times OSp(1|32)$,
hep-th/0306269

\bibitem{btz} M. Banados, R. Troncoso and J. Zanelli, ``Higher dimensional 
Chern-Simons supergravity'', Phys. Rev. D 54 (1996) 2605 and gr-qc/9601003

\bibitem{tz1} R. Troncoso and J. Zanelli, ``New gauge supergravity in 
seven and eleven dimensions'', Phys. Rev. D58 (1998) 101703
 and hep-th/9710180

\bibitem{tz2} R. Troncoso and J. Zanelli, ``Chern-Simons supergravities
with off-shell superalgebras'',  hep-th/9902003

\bibitem{zanelli} J. Zanelli, ``Chern-Simons gravity: from 2+1 to 2n+1 
dimensions'', Braz. J. Phys. 30 (2000) 251 and hep-th/0010049

\bibitem{horava} P. Horava, ``M-theory as a holographic theory,''
Phys. Rev. D59 (1999) 046004  and hep-th/9712130

\bibitem{lanczos} C. Lanczos, Ann. Math. 39 (1938) 842

\bibitem{lovelock} D. Lovelock, J. Math. Phys. 12 (1971) 498

\bibitem{acotz} R. Aros, M. Contreras, R. Olea, R. Troncoso, J. Zanelli, "Conserved charges 
for gravity with locally AdS asymptotics," Phys. Rev. Lett. 84 (2000) 1647 and 
gr-qc/9909015; R. Aros, M. Contreras, R. Olea, R. Troncoso, J. Zanelli,
"Conserved charges for even dimensional asymptotically AdS gravity theories," 
Phys.Rev. D62 (2000) 044002 and hep-th/9912045; R. Olea, "Mass, angular momentum 
and thermodynamics in four-dimensional Kerr-AdS black holes," JHEP 0506 (2005) 023
and hep-th/0504233.

\bibitem{dn} B. de Wit and H. Nicolai, 
"On the relation between d=4 and d=11 supergravity," Nucl. Phys. B 243 (1984) 91;
B. de Wit, H. Nicolai and N.P. Warner, "The embedding of gauged N=8 supergravity into
d=11 supergravity," Nucl. Phys. B255 (1985) 29;
B. de Wit and H. Nicolai, "Hidden symmetry in d=11 supergravity,"
Phys. Lett. B155 (1985) 47 and "D=11 supergravity with local SU(8) invariance," Nucl. Phys.
B274 (1986) 363 and "The consistency of the $S^7$ truncation of d=11 supergravity,"
Nucl. Phys. B281 (1987) 211.

\bibitem{nvv} H. Nastase, D. Vaman and P. van Nieuwenhuizen, 
"Consistent nonlinear K K reduction of 11-d supergravity on AdS(7) x S(4) and 
selfduality in odd dimensions," Phys. Lett. B469 (1999) 96 and hep-th/9905075; and
"Consistency of the 
AdS(7) x S(4) reduction and the origin of selfduality in odd dimensions," 
Nucl. Phys. B581 (2000) 179 and hep-th/9911238.

\bibitem{chamseddine} A.H. Chamseddine, ``Topological gauge theory of 
gravity in five and all odd dimensions'' Phys. Lett. B 233 (1989) 291;
``Topological gravity and supergravity in various dimensions'', Nucl. 
Phys. B 346 (1990) 213

\bibitem{cjs} E. Cremmer, B. Julia and J. Scherk, ``Supergravity theory 
in 11 dimensions'', Phys. Lett. B 76 (1978) 409

\bibitem{df} R. D'Auria and P. Fre, ``Geometric supergravity in d=11 and 
its hidden supergroup'', Nucl. Phys. B 201 (1982) 101

\bibitem{cfgpv2} L. Castellani, P. Fre, F. Giani, K. Pilch and P. van 
Nieuwenhuizen, ``Beyond d=11 supergravity and Cartan integrable systems,''
Phys. Rev. D26 (1982) 1481








\end{thebibliography}
\end{document}